\documentclass[11pt,twoside]{article}

\usepackage{asp2006}
\usepackage{epsf}
\usepackage{psfig}
\usepackage{lscape}

\begin{document}
\title{DD Mon and XY~UMa: CCD Photometry and modelling of two close binary systems with
solar-type components}
\author{K. Gazeas$^{1,2}$, A. Liakos$^{2}$ and P. Niarchos$^{2}$}
\affil{$^{1}$Harvard Smithsonian Center for Astrophysics, 60
Garden Street, Cambridge, MA 02138, USA\\
$^{2}$Department of Astrophysics, Astronomy and Mechanics,
National and Kapodistrian University of Athens, GR 157 84
Zografos, Athens, Greece}

\begin{abstract}
We present our ground-based CCD observations of the close binary
systems DD~Mon and XY~UMa in B, V, R and I bands. The light curves
are analyzed using the Wilson-Devinney code (W-D) for the
derivation of the geometric and photometric elements of the
systems. We compare the methods of photometric and spectroscopic
mass ratio determination in these binaries, as a function of all
typical difficulties, which arise during the analysis of such
systems (light curve asymmetries, third light etc). Finally, a new
spot model is suggested for the eclipsing system XY~UMa, which
belongs to the RS~CVn type of active binaries.
\end{abstract}

\section{Observations and light curve analysis}
The B, V, R and I-band observations of both systems were carried
out by means of CCD differential photometry. DD~Mon was observed
on March 10-13, 2005 with the 1.22m Cassegrain reflector at the
Kryoneri Astronomical Station of the National Observatory of
Athens, Greece, while XY~UMa was observed on December 1, 4, 5 and
8, 2006 with the 0.40m Cassegrain reflector at the University of
Athens Observatory, Greece. The light curves were analyzed by
using the PHOEBE 0.29d software \citep{PZ05}, which utilizes the
W-D code \citep{WD71,W79}.

For the photometric light curve analysis, we performed a q-search
on both systems in modes 2, 4, 5, for a rough estimation of the
photometric mass ratio ($q_{ph}$). We chose the range of
${0<q<1}$, and the best value (minimum sum of the square residuals
- $\chi^{2}$) of $q_{ph}$ was later used for the final solution
(see Tables 2,~3, Fig.1). In both cases, the photometric mass
ratios ($q_{ph}=0.575$ for DD~Mon and $q_{ph}=0.484$ for XY~UMa)
were found to be smaller than the one obtained spectroscopically,
which are $q_{sp} = 0.670(19)$ for DD~Mon \citep{P09} and $q_{sp}
= 0.70$ for XY~UMa \citep{P07}.

The geometric and physical elements for both systems, together
with the above radial velocities, are used to compute the absolute
physical parameters (radii, masses and luminosity) in solar units
(Table 1). In this table, we also include the solutions, derived
with the use of the spectroscopically determined mass ratio,
utilizing the (directly observed) radial velocities K1 and K2 as
fixed parameters.

%-------------------Beginning of Table 1 ------------------------------
\begin{table*}
\caption{Comparison of the absolute elements of DD~Mon and XY~UMa,
obtained with photometrically and spectroscopically calculated
mass ratio.}
\begin{flushleft}
\begin{small}
\scalebox{0.95}{
\begin{tabular}{lcccccc}

\hline \hline
\textbf{Elem.}&    \multicolumn{3}{c}{\textbf{DD Mon}}   &   \multicolumn{3}{c}{\textbf{XY UMa}}    \\
              &$q_{ph}=0.575$&$q_{sp}=0.670$&$diff.~(\%)$&$q_{ph}=0.484$&$q_{sp}=0.610$&$diff.~(\%$)\\
\hline
$M_{1}$       &    1.94(9)   &    1.39(7)   &    32.9    &    1.85(3)   &    1.13(8)   &    47.8    \\
$M_{2}$       &    1.12 (6)  &    0.93(6)   &    17.8    &    0.89(2)   &    0.69(4)   &    26.1    \\
$R_{1}$       &    1.62(3)   &    1.44(3)   &    12.1    &    1.37(1)   &    1.14(3)   &    17.8    \\
$R_{2}$       &    1.37(3)   &    1.29(3)   &    5.6     &    0.73(1)   &    0.68(1)   &    6.8     \\
$L_{1}$       &    3.54(13)  &    2.78(11)  &    24.1    &    1.32(2)   &    0.92(5)   &    35.3    \\
$L_{2}$       &    1.21(5)   &    1.08(5)   &    11.6    &    0.098(3)  &    0.093(4)  &    5.2     \\
$M_{bol1}$    &    3.36(4)   &    3.63(4)   &    7.5     &    4.44(1)   &    4.83(5)   &    8.4     \\
$M_{bol2}$    &    4.53(5)   &    4.66(5)   &    2.7     &    7.25(3)   &    7.31(5)   &    0.8     \\
\hline \hline

\end{tabular}}
\end{small}
\end{flushleft}
\begin{small}
\end{small}
\end{table*}
%-------------------End of Table 1 ------------------------------------

%-------------------Beginning of Table 2 ------------------------------

\begin{table*}

\caption{The parameters of DD Mon derived from the LCs solution.}
\scalebox{0.6}{
\begin{tabular}{lcccc|cccc}
\hline \hline
\textbf{Parameter}          &       \multicolumn{4}{c}{mode 5, adjusted q}      &            \multicolumn{4}{c}{mode 5, fixed q}     \\
\hline
$\varphi_{0}$               &            \multicolumn{4}{c}{-0.0004(1)}         &                \multicolumn{4}{c}{-0.0004(1)}      \\
$q~(m_{2}/m_{1})$           &            \multicolumn{4}{c}{0.575(3)}           &           \multicolumn{4}{c}{0.670}                \\
$i$ [deg]                   &            \multicolumn{4}{c}{78.2}               &               \multicolumn{4}{c}{77.2(1)}          \\
$T_1$$^{*}$[K], $T_2$ [K]   &          \multicolumn{4}{c}{6250, 5202(4)}        &           \multicolumn{4}{c}{6250, 5195(4)}        \\
$A_1$$^*$=$A_2$$^*$         &            \multicolumn{4}{c}{0.5}                &                   \multicolumn{4}{c}{0.5}          \\
$g_1$$^*$=$g_2$$^*$         &           \multicolumn{4}{c}{0.32}                &               \multicolumn{4}{c}{0.32}             \\
$\Omega_{1}$, $\Omega_{2}$  &        \multicolumn{4}{c}{3.230(10), 3.037}       &       \multicolumn{4}{c}{3.410(4), 3.190}          \\
                            &     $B$    &     $V$    &     $R$    &     $I$    &     $B$    &     $V$    &     $R$    &     $I$     \\
$L_{1}/L_{T}$               & 0.757(2)   &  0.711(2)  &  0.685(3)  &   0.658(3) &   0.730(2) &   0.685(2) &    0.659(2)&  0.633(3)   \\
$L_{2}/L_{T}$               &0.1959(3)   &  0.2242(4) & 0.2438(7)  &  0.2636(7) &   0.2133(3)&  0.2443(4) &   0.2657(5)&  0.2874(8)  \\
$L_{3}/L_{T}$               &0.047(2)    & 0.065(2)   &   0.072(3) &   0.079(4) &   0.056(2) &  0.071(2)  &   0.075(3) &  0.080(4)   \\
$X_{1}$, $X_{2}$            &0.677, 0.832&0.547, 0.689&0.468, 0.595&0.390, 0.501&0.677, 0.832&0.547, 0.690&0.468, 0.595&0.390, 0.502 \\
\hline
                            &   $pole$   &  $point$   &   $side$   &   $back$   &   $pole$   &  $point$   &   $side$   &   $back$    \\
\hline
$r_{1}$                     &  0.3702(1) & 0.4292(24) & 0.3867(12) & 0.4055(15) & 0.3598(5)  & 0.4178(12) & 0.3752(6)  &  0.39448)   \\
$r_{2}$                     &  0.3123(5) & 0.4452(8)  & 0.3263(5)  & 0.3586(5)  & 0.3231(6)  & 0.4589(10) & 0.3380(6)  &  0.3700(8)  \\
\hline
$\chi^{2}$                  &             \multicolumn{4}{c}{0.4221}            &                \multicolumn{4}{c}{0.4327}          \\
\hline\hline
                            & \multicolumn{4}{c}{\textit{$^*$assumed}, \textit{$L_{T}=L_{1}+L_{2}+L_{3}$}}&\multicolumn{4}{c}{}      \\
\end{tabular}}
\end{table*}

%-------------------End of Table 2 ------------------------------------

%-------------------Beginning of Table 3 ------------------------------

\begin{table*}

\caption{The parameters of XY UMa derived from the LCs solution.}
\scalebox{0.57}{
\begin{tabular}{lcccc|cccc}
\hline \hline
\textbf{Parameter}          &          \multicolumn{4}{c}{mode 5, adjusted q}   &       \multicolumn{4}{c}{mode 5, fixed q}          \\
\hline
$\varphi_{0}$               &            \multicolumn{4}{c}{-0.0024(1)}         &           \multicolumn{4}{c}{-0.0024(1)}           \\
$q~(m_{2}/m_{1})$           &            \multicolumn{4}{c}{0.484(2)}           &               \multicolumn{4}{c}{0.610}            \\
$i$ [deg]                   &            \multicolumn{4}{c}{80.6(1)}            &               \multicolumn{4}{c}{77.6(1)}          \\
$T_1$$^{*}$[K], $T_2$ [K]   &          \multicolumn{4}{c}{5310, 3889(6)}        &           \multicolumn{4}{c}{5310, 3806(6)}        \\
$A_1$$^*$=$A_2$$^*$         &            \multicolumn{4}{c}{0.5}                &            \multicolumn{4}{c}{0.5}                 \\
$g_1$$^*$=$g_2$$^*$         &           \multicolumn{4}{c}{0.32}                &           \multicolumn{4}{c}{0.32}                 \\
$\Omega_{1}$, $\Omega_{2}$  &        \multicolumn{4}{c}{3.190(5), 3.670(10)}    &       \multicolumn{4}{c}{3.425(3), 3.983(10)}      \\
                            &     $B$    &     $V$    &     $R$    &    $I$     &     $B$    &     $V$    &     $R$    &   $I$       \\
$L_{1}/L_{T}$               &   0.939(3) &  0.940(2)  & 0.928(2)   &  0.865(2)  &   0.947(3) &   0.940(3) &   0.924(2) &  0.869(2)   \\
$L_{2}/L_{T}$               &   0.0317(1)&  0.0475(1) & 0.0607(1)  &  0.0721(1) &   0.0343(1)&  0.0523(1) &  0.0678(1) &  0.0833(1)  \\
$L_{3}/L_{T}$               &   0.030(3) &  0.013(2)  &   0.012(1) &   0.063(1) &   0.019(2) &  0.008(2)  &  0.008(2)  &  0.047(1)   \\
$X_{1}$, $X_{2}$            &0.849, 0.823&0.787, 0.796&0.720, 0.770&0.633, 0.683&0.849, 0.828&0.787, 0.800&0.720, 0.770&0.633, 0.682 \\
\hline
                            &   $pole$   &  $point$   &   $side$   &   $back$   &   $pole$   &  $point$   &   $side$   &   $back$    \\
\hline
$r_{1}$                     & 0.3650(7)  & 0.4082(12) & 0.3795(8)  & 0.3935(9)  & 0.3503(4)  & 0.3937(6)  & 0.3635(4)  & 0.3785(5)   \\
$r_{2}$                     & 0.1994(10) & 0.2086(13) & 0.2019(11) & 0.2068(12) & 0.2135(7)  & 0.2234(9)  & 0.2163(8)  & 0.2214(8)   \\
\hline
\textbf{Spot Parameter}     & $star~No~1$&  $spot~1$  &  $spot~2$  &            & $star~No~1$&  $spot~1$  & $spot~2$   &             \\
\hline
$co-latitude$ [deg]         &            &  82.3(4)   &  81.6(2)   &            &            &  82.4(3)   &  82.2(3)   &             \\
$longitude$ [deg]           &            &  192.6(3)  &  139.2(1)  &            &            &  190.6(3)  & 133.5(1)   &             \\
$radius$ [deg]              &            &  10.9(1)   &  11.4(2)   &            &            &  10.9(1)   &  11.4(2)   &             \\
$temp.~factor$              &            &  0.759(4)  & 0.765(4)   &            &            & 0.722(3)   &  0.759(4)  &             \\
\hline
$\chi^{2}$                  &              \multicolumn{4}{c}{0.0890}           &              \multicolumn{4}{c}{0.1559}            \\
\hline\hline
                            & \multicolumn{4}{c}{\textit{$^*$assumed}, \textit{$L_{T}=L_{1}+L_{2}+L_{3}$}}&\multicolumn{4}{c}{}      \\
\end{tabular}}
\end{table*}

%-------------------End of Table 2 ------------------------------------

\begin{figure}[]
\plottwo{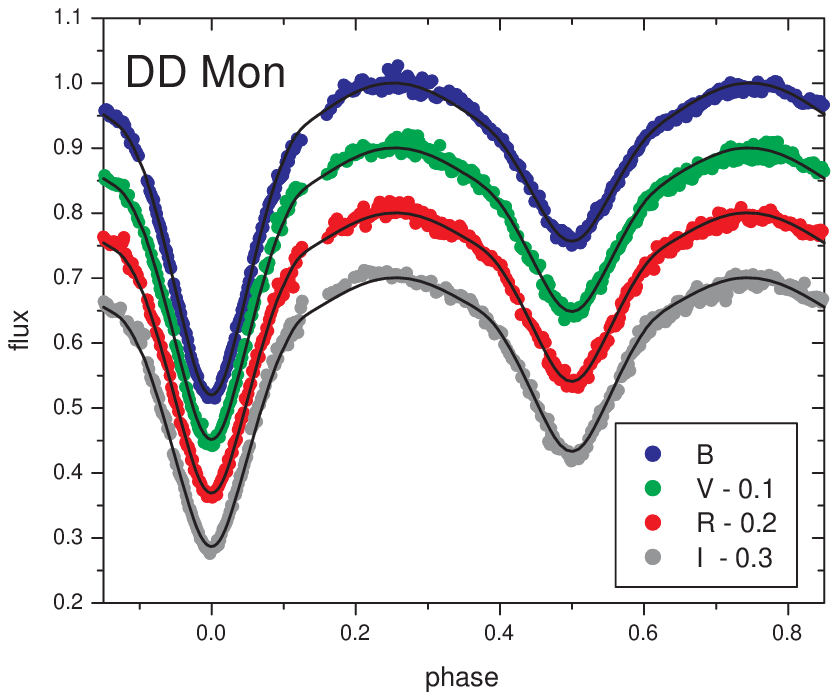}{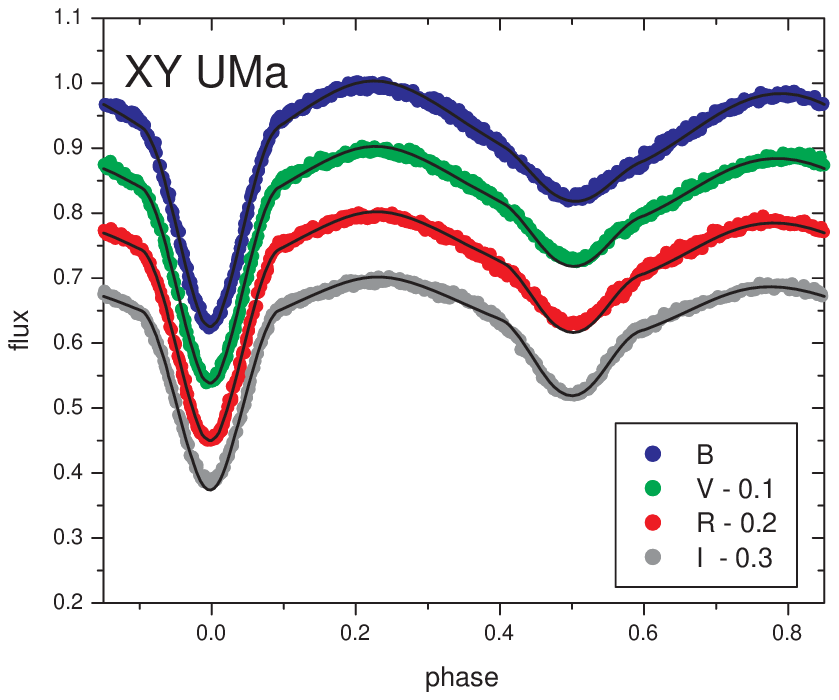}

\caption{The observed and synthetic light curves of DD Mon and XY
UMa.}
\end{figure}

\section{Summary and conclusions}
The determination of the orbit of the secondary component is a
major issue on near contact and detached systems, where the
temperature and luminosity difference between the two components
is large. Such a difficulty affects the mass ratio determination,
where the proximity and eclipse effects play an additional role
\citep{ND03,VHW85}.

The radial velocities measured on XY~UMa components are affected
by the presence of this effect. Therefore, the hotter area of the
secondary is observed in smaller wavelengths and therefore smaller
radial velocity is measured. This effect is also noticed by
\citet{P09}, who observed XY~UMa on the Mg triplet region of the
spectrum and found a mass ratio of $q_{sp} = 0.70$, which is
significantly larger than the the value of $q_{sp} = 0.61$
\citep{PJ98}, found by infrared spectroscopy.

In our study we found that the photometric mass ratios for both
systems were significantly smaller than the spectroscopic ones,
which might be a result of the reflection effect. The geometric
and physical elements describe very well the systems, using either
the photometric or the spectroscopic mass ratio and the two
theoretical models are distinguished only by the residual levels.
However, the calculated absolute elements resulted in larger
masses, radii and luminosities for both components. We conclude
that the solutions based only on photometric data may hide such
effects, giving solutions which might be even by 50\% different
than those expected.

\end{document}